# Unconventional topological edge states in one-dimensional gapless systems stemming from nonisolated hypersurface singularities

Hongwei Jia[1,2,#,*], Jing Hu[2,3,#,†], Ruo-Yang Zhang[2], Yixin Xiao[2], Dongyang Wang[4], Mudi Wang[2], Shaojie Ma[5], Xiaoping Ouyang[6], Yifei Zhu[7], C. T. Chan[2,‡].

[1]*Institute of Precision Optical Engineering, School of Physics Science and Engineering, Tongji University, Shanghai 200092, China.*

[2]*Department of Physics, Hong Kong University of Science and Technology, Hong Kong, China.*

[3]*School of Materials Science and Engineering, Shanghai University, Shanghai, 200444 China.*

[4]*Optoelectronics Research Centre, University of Southampton, Southampton SO17 1BJ, United Kingdom.*

[5]*Shanghai Engineering Research Centre of Ultra Precision Optical Manufacturing, Department of Optical Science and Engineering, School of Information Science and Technology, Fudan University, Shanghai 200433, China.*

[6]*School of Materials Science and Engineering, Xiangtan University, Xiangtan, China.*

[7]*Department of Mathematics, Southern University of Science and Technology, Shenzhen, China.*

[#]These authors contributed equally to this work

[*]jiahongwei7133@gmail.com; [†]huj@shu.edu.cn; [‡]phchan@ust.hk.

**Abstract**: Topologically protected edge states have been extensively studied in systems characterized by the topological invariants in band gaps (also called line gaps). In this study, we unveil a whole new form of edge states that transcends the established paradigms of band-gap topology. In contrast to the traditional stable edge states in topological insulators with specific band gaps, the one-dimensional systems we investigate are inherently gapless with the Brillouin zones being mapped to the loops encircling hypersurface singularities in a higher-dimensional space with parity-time symmetry. These hypersurface singularities are nonisolated degeneracies embedded entirely on exceptional surfaces, rendering the energy gaps in our systems inevitably closed at the intersections of the Brillouin zone loop and the exceptional surfaces. Unexpectedly, such gapless systems still afford topologically protected edge states at system boundaries, challenging the conventional understanding based on band gaps. To elucidate the existence of these edge states in the absence of a band-gap-based invariant, we propose a theoretical framework based on eigen-frame rotation and deformation that incorporates non-Bloch band theory. Finally, we experimentally demonstrate this new form of topological edge states with nonreciprocal circuits for the first time. Our work constitutes a major advance that extends topological edge states from gapped phases to gapless phases, offering new insights into topological phenomena.

**Main:** The theory of band topology has emerged as a robust framework for understanding diverse and intriguing physical phenomena [1-24]. Central to this theory is the physical concept called bulk-edge correspondence, occupying a pivotal role in predicting the stable edge states based on the bulk topology of the system [15-24]. The topologically protected edge states have been extensively explored in conventional Hermitian systems, including topological insulators [15-17], Weyl/Dirac semimetals [18-22], and nodal line metals [23,24]. In recent years, non-Hermitian physics has gained increasing significance, and the notion of bulk-edge correspondence has undergone a remarkable extension into the domain of non-Hermitian systems, encompassing a generalized framework that duly accounts for non-Hermitian skin effects [25-30]. Stable band degeneracies, protected by system symmetries, can be viewed as topological defects of eigenenergies or eigenstates in momentum space [31]. The exceptional points (EPs), which refer to coalescences of both eigenvalues and eigenstates exclusively from non-Hermiticity, are generating growing research interest [32-39]. EPs can further aggregate into surfaces, dubbed exceptional surfaces (ESs), under the protection of parity-time (*PT*) symmetry or chiral symmetry [40-46]. It has been recognized that ESs correspond to the singular hypersurfaces in momentum space. This has inspired us to utilize catastrophe theory [47] to comprehend and predict various hypersurface singularities embedded on ESs as gapless degeneracy structures, such as transversal intersections [43], cusps [44,46], and exotic swallowtail catastrophes [45], etc. Despite the importance of the topological edge states, its association with hypersurface singularities in non-Hermitian systems has hitherto remained elusive. This stems from the prevailing belief that a well-defined band gap is a prerequisite for the existence of band topology capable of protecting boundary modes. Nevertheless, it remains counter-intuitive that boundary modes can emerge in systems that lack band gaps, where it is impossible to define topological invariants based on band gaps.

This study delves into this uncharted but important territory, where we discover a new form of topological edge states beyond the established paradigms of band gap topology. Hypersurface singularities in band structures, which are a form of gapless degeneracy feature, provide ideal platforms for investigating topological phenomena in gapless phases. We thus explore the edge states associated with a structurally rich hypersurface singularity, i.e., swallowtail catastrophe [45]. Since the swallowtail hosts various types of degeneracy lines that locate entirely on ESs, closed loops encircling such degeneracy lines inevitably intersect ESs. Consequently, when these loops are mapped onto the Brillouin zones (BZ) of one-dimensional (1D) periodic systems, EPs emerge, at which the gap closes. While such 1D systems are essentially gapless phases, they still host topologically protected edge states localized at system boundaries. As an obvious distinction from conventional gapped systems, band-gap based topological invariants cannot be defined for such gapless systems, we propose a theoretical framework of eigen-frame rotation and deformation incorporating the non-Bloch band theory to understand this new form of topological edge states [27,48,49]. To substantiate our findings, we implement these 1D systems with topological circuits, providing the first experimental demonstration of this new form of topological edge states.

To grasp a direct visualization of the novel form of edge states that are supported by topological gapless systems, we start with a sketched comparison against the conventional form. Topologically protected edge states are generally discussed upon topological insulators. A 1D topological insulator can be generically constructed by embedding its BZ (e.g., a 1D homotopy loop [31]) into a higher dimensional space and to encompass an isolated degeneracy, as illustrated in FIG. 1(a). Topological invariants can be defined for a specific band gap by examining the geometric phases of all the bands below that gap. This approach is applicable to non-Hermitian systems, provided that a line gap persists [see FIG. 1(b)]. The eigenvalues of the conventional edge states commonly reside inside the gaps between the bulk continuums in the complex projected spectrum, as displayed in FIG. 1(c). The existence of these edge states can be traced to this gap-based topological invariant. In contrast, the unconventional edge states discovered in this work are associated with a fundamentally different form of degeneracies, i.e., the hypersurface singularities embedded entirely on ESs, as shown in FIG. 1(d). The 1D gapless system can still be constructed by engineering the BZ to enclose the hypersurface singularities. However, such a loop inevitably intersects the degenerate ESs and is thus an intersection homotopy loop [50]. This leads to a critical fact that the system is inherently gapless, as shown in FIG. 1(e). Due to the presence of exceptional degeneracies on the BZ, the bulk continuums are no longer isolated from each other but are overlapped in projected spectrum, resulting in the elimination of gaps, as shown in FIG. 1(f). Focusing on this important and unexplored class of systems, we theoretically and experimentally demonstrate that a new form of edge states can stably exist in such gapless phases with eigenvalues away from the overlapped bulk continuum, as sketched in FIG. 1(f). Notably, this characteristic distinguishes the proposed edge states from bound states in the continuum (BICs) [51]. While the eigenvalues of BICs lie within the continuous spectrum of bulk or skin modes, they remain decoupled from these modes. In contrast, the novel edge state lies outside the bulk or skin continuum in the complex plane. Due to the lack of line gaps in the system and the Hamiltonian matrix becoming defective at certain $k$-points, conventional topological invariants, such as the Berry phase, cannot be defined.

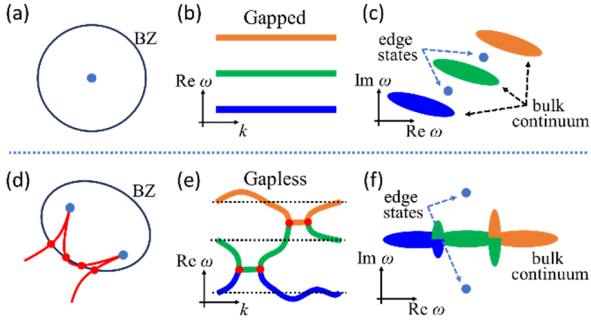

FIG. 1. Sketch of the comparison between conventional edge states and unconventional edge states in 1D topological phases. (a) The 1D BZ of a topological insulator encircles an isolated degeneracy in the higher parameter space. (b) The BZ of the system does not intersect any degeneracy point, and thus the eigenvalues are all gapped over the BZ. Here $k$ is the Block wave vector of the 1D system, and $\omega$ is the eigenvalue of the 1D periodic system. (c) The projected complex spectrum of the 1D topological insulator. The eigenvalues of the topological edge modes reside in the line gaps between the bulk continuums. (d-f) Similar to (a-c) but for a system with BZ encircling nonisolated hypersurface singularities.

To construct such 1D systems, we first introduce the structurally rich hypersurface singularity, i.e., swallowtail catastrophe, with a simple three-state non-Hermitian Hamiltonian

$$H(f_x, f_y, f_z) = \begin{bmatrix} 1 & -f_x & -f_y \\ f_x & 0 & -f_z \\ f_y & -f_z & 0 \end{bmatrix} \quad (1)$$

in the auxiliary 3D **f**-space (represented by real parameters $f_x$, $f_y$ and $f_z$), which manifests $PT$ symmetry and an additional $\eta$-pseudo-Hermitian symmetry [52] $\eta H \eta^{-1} = H^\dagger$ (the metric operator takes the Minkowski metric $\eta = \mathrm{diag}(-1,1,1)$ [53]). The degeneracy structure is displayed in FIG. 2(a), with the purple surfaces denoting ESs. There exist three seemingly disparate yet topologically interconnected types of degeneracy lines: a pair of exceptional lines of order three (EL3, blue lines) located at the cusps of ESs, a nondefective intersection line (NIL, upper black line) where ESs intersect transversally, and a nodal line (NL, lower black line) isolated from ESs. These three distinct types of degeneracy lines are stably connected at the meeting point (MP, yellow star) [45]. Their nearby dispersions are shown in **End Matter** (FIG. 4). It is seen that the swallowtail hosts various types of degeneracy lines embedded on ESs, and is therefore an ideal parent platform for generating 1D gapless Hamiltonian exhibiting diverse non-Hermitian topologies. The 1D gapless systems can then be designed by selecting closed loops that encircle different degeneracy lines, each loop represents the Brillouin zone of a one-dimensional non-Hermitian system, i.e., expressing $f_{x,y,z}$ with real functions of $\sin k$ or $\cos k$ [$f_{x,y,z}(\sin k, \cos k)$]. A 1D Hamiltonian can thus be obtained $H(k)$, with the parameter $k$ interpreted as the 1D Bloch wave vector and its interval $[0, 2\pi]$ constitutes the BZ of the periodic system.

The corresponding schematic diagrams of the 1D models in real space, with open boundary conditions (OBC) or periodic boundary conditions (PBC), are depicted pictorially in the upper and lower panels of FIG. 2(b), respectively. The dashed blocks in the panels enclose two unit cells along with their schematic connecting hoppings between the orbitals, while the internal structure is detailed in FIG. 2(c). Three orbitals **A**, **B**, and **C** are required within each unit cell to realize the three-state Hamiltonian. These hoppings can be reciprocal or non-reciprocal, as labelled by orange and purple arrows, respectively. Such 1D systems can be experimentally implemented using topological circuits, with the admittance bands $j$ corresponding to the energy bands derived from the Hamiltonian in a specific tight-binding model [30]. Circuit systems offer a distinct advantage over other platforms as they can conveniently implement non-reciprocal hoppings in a complex system, owing to the availability of various active circuit elements, e.g., operational amplifier (OpAmp). The 1D lattice circuits in our experiment consist of eleven unit cells, which are designed based on the real space Hamiltonian with hoppings illustrated in FIG. 2(c). This schematic diagram displays two unit cells and the corresponding circuit design is shown in FIG. 2(d). In addition to the conventional elements (e.g., capacitors, inductors and resistors),

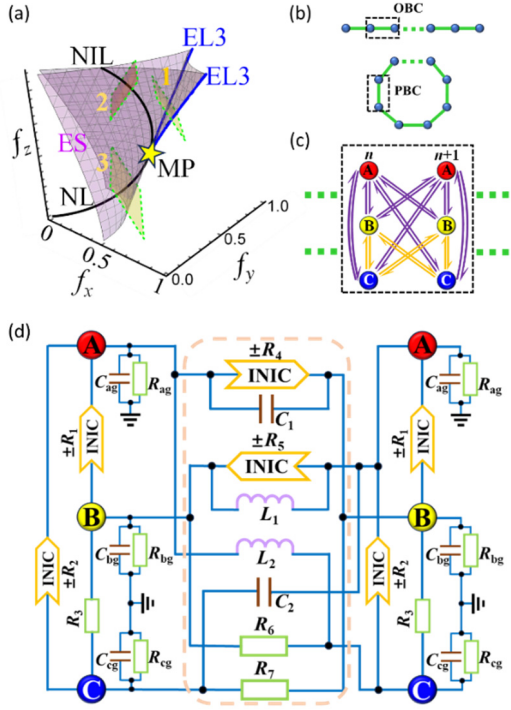

FIG. 2. Implementation of 1D gapless systems based on the non-Hermitian swallowtail catastrophe. (a) Geometric structure of the swallowtail in 3D **f**-space, which has three different types of degeneracy lines, i.e., a pair of EL3s at cusps of ESs (the blue lines), an NIL at the transversal intersection of ESs (upper half of the black line), and an NL isolated from the ESs (lower half of the black line). These degeneracy lines are stably connected at the same point, i.e., the MP (yellow star). (b) Schematic diagram of the periodic system. The upper panel and the lower panel show the system under PBC and OBC, respectively. (c) Internal structure of the dashed block in panel (b). The dashed block in (b) encloses two unit-cells along with hoppings connecting any two sublattices **A**, **B** and **C**. The bi-oriented arrows represent forward and backward hoppings between sublattices. Orange arrows indicate Hermitian hoppings, where the forward and backward hopping parameters are complex conjugates. In contrast, purple arrows represent nonreciprocal non-Hermitian hoppings, where the hopping parameters are negatively conjugated. (d) The designed circuit schematic of two unit cells [corresponding to panel (c)]. The reciprocal hopping parameters are simply realized with pure resistors ($R$). The non-reciprocal hopping parameters are implemented using capacitors ($C$), inductors ($L$) and INIC ($\pm R_n$) circuit accordingly.

negative impedance converter through current inversion (INIC)

circuits are utilized to realize the non-reciprocal hoppings. By adjusting the hopping parameters of the 1D lattice models, their respective BZs correspond to closed loops that encircle different degeneracy lines. Through the measurement of the voltage response at each node to a local current input in each model, one can obtain the admittance band structures, encompassing both continuous states and edge states. More details on the experiments are provided in Section 1 of Supplemental Materials.

The swallowtail reveals several interesting topological interconnections, e.g., the pair of EL3s is topologically equivalent to the NL and NIL together [45]. Therefore, our systems also allow us to examine the relationships of edge states associated with different types of singular lines. We start by examining a 1D system obtained by mapping the BZ to the loop $l_a$ encircling the NL and NIL together, as shown in FIG. 3(a). Notably, a 1D system with BZ encircling either a single NL or a single NIL hosts a single pair of edge states, we show relevant data in **End Matter** (FIG. 5). Here the BZ of the system is designed to intersect ESs twice for the convenience of experimental implementation. Therefore, $j_2$ and $j_3$ undergo gap closing twice on the BZ, as shown in FIG. 3(b) (the eigenvalues $j_{1\sim3}$, colored in black, blue and red, are sorted with their real-valued magnitudes in exact phase sectors). Despite this gap closing, two distinct pairs of edge states are still observable (labelled by 1, 2 and 3, 4), as illustrated by the experimental and theoretical outcomes in FIG. 3(c). Skin modes also evident, as the complex eigenvalues under OBC (brown) collapse into open arcs within the region enclosed by the PBC (olive) spectrum. The experimental results of the field distributions for the four edge states are shown in FIG. 3(d). Obviously, the fields are confined near the two edges. Although the eigenvalues of the edge states reside in the gap between the continuum of $j_1$ and the merged continuum of $j_{2,3}$, the emergence of edge states is indeed independent of this gap. This features that the edge states are stable against gap closing (by deforming the BZ to intersect different ESs) are distinctive from that of the conventional edge states, and thus transcend the understanding based on the topological invariant defined upon band gaps (e.g., Berry phase).

To understand the topological mechanism, we propose a theoretical framework based on eigenframe deformation and rotation processes that incorporates the non-Hermitian skin effect (Sections 2-3 in Supplemental Materials). Such a theory well addresses the gap closing at EPs, because the topological characterization is independent of how the conjugate bands (i.e.,

$j_2$ and $j_3$) are ordered in the broken phase sector (see Sections 6 of Supplemental Materials). We investigate the accumulated angle of the eigenstates ($\varphi_1$, $\varphi_2$ and $\varphi_3$, the same colors as $j_{1\sim3}$) via the eigenframe deformation and rotation processes [defined by $\theta_m = \arccos(\varphi_m^T \cdot \varphi_{m,i})$, $m$=1, 2, 3, $\varphi_{m,i}$ denotes the initial state at the starting point] along the generalized Brillouin zone (GBZ, $k'$ is the non-Bloch wave vector), as displayed in FIG. 3(e). This evolution process indicates a multiband topology, because both $\varphi_1$ and $\varphi_3$ experience quantized $\pi$ accumulated angles, while $\varphi_2$ accumulates a trivial angle of zero. Such a quantization behavior results from PT symmetry. This explains why there exist two pairs of edge states away from the continuums. The imaginary parts just serve as intermediate processes as the initial and final $\theta$ are all real, despite that the imaginary parts of $\theta$ are nonzero in the broken phase sector where the eigenstates are complex. Detailed eigenframe evolution is shown in FIG. S9(f) in Supplemental Materials indicating $\hat{O}\varphi_1 = -\varphi_1$, $\hat{O}\varphi_3 = -\varphi_3$ and $\hat{O}\varphi_2 = \varphi_2$ ($\hat{O}$ denotes the evolution operator along the GBZ) due to their respective accumulated angles. This nontrivial topology represented by the evolution of eigenframe ensures the existence of the two pairs of edge states. It is obvious that the gap-based topological invariant is inapplicable here (e.g., ill-defined Berry connection at EPs because the matrix is defective). We provide detailed discussions on the challenges faced by the band gap theory in understanding this novel form of edge states in Section 7 of Supplemental Materials.

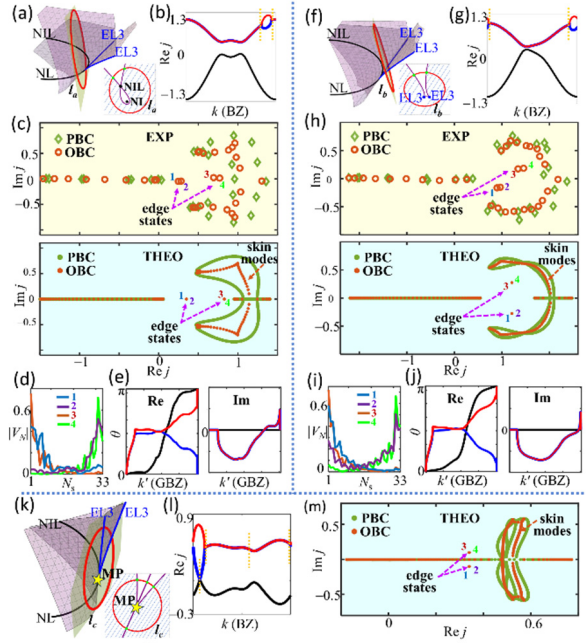

FIG. 3. Observation of topological edge states in 1D gapless systems with BZs encircling the NIL and NL together and the double EL3s. (a) The BZ of the 1D system in **f**-space that encircles the NIL and NL together (red loop, $l_a$). The inset shows the plane that the loop resides on, with shaded and unshaded regions being broken and exact phase domains, respectively. The red dots show the intersections between $l_a$ and ESs (b) Band dispersions (real part) on the BZ of the system (or on $l_a$). The BZ has two EPs (yellow dashed lines) where $j_2$ and $j_3$ are degenerate. (c) Experimental (upper) and theoretical (lower) results of the complex spectra of the system. Two pairs (labelled by 1, 2 and 3, 4) of topological edge states with eigenvalues being real are shown by the purple arrows. (d) Experimental results of the field distributions for the two pairs of the edge states [i.e., edge states 1, 2 and edge states 3, 4 in panel (c)]. Here, the voltage amplitude at each sublattice site represents the output response to the input current signal associated with specific eigenvalues, and the voltage distribution corresponds precisely to the experimental result of the field profile of the edge states. These fields exhibit clear confinement near the edges. (e) Accumulated angles for the eigenstates $\varphi_1$, $\varphi_2$ and $\varphi_3$ via the eigenframe rotation and deformation process as the non-Bloch wave vector varies along the GBZ. (f-j) Similar to (a-e) but for BZ encircling the double EL3s (i.e., $l_b$). (k-m) Similar to (a-c) but for BZ encircling the MP, which intersects the ESs formed by $j_{1,2}$ and $j_{2,3}$ (i.e., $l_c$).

We next move to the edge states associated with loop encircling the double EL3s [$l_b$ in FIG. 3(f)]. The loop $l_b$ inevitably cuts through the ESs twice (see inset), and correspondingly, the bands $j_2$ and $j_3$ experience eigenvalue coalescence twice, as shown by the dispersion in FIG. 3(g). The complex spectra under PBC and OBC are displayed in FIG. 3(h), corresponding to experimental and theoretical results, respectively. Similar to the former system with BZ being $l_a$, the system with BZ being $l_b$ also holds two pairs of edge states (labelled 1, 2 and 3, 4), with their complex eigenvalues residing outside the bulk continuum and conjugate to each other. The corresponding field distributions, obtained via experimental measurements, are shown in FIG. 3(i). We need to check whether the accumulated angles of the eigenframe are identical to those for the system encompassing the NL and the NIL together. As can be observed in FIG. 3(j), $\varphi_1$ and $\varphi_3$ accumulate $\pi$, and $\varphi_2$ accumulates zero, which is the same as FIG. 3(e) except for slight differences in the intermediate process, $l_a$ and $l_b$ are along different trajectories. The accumulated angles result in $\hat{O}\varphi_1 = -\varphi_1$, $\hat{O}\varphi_3 = -\varphi_3$, and $\hat{O}\varphi_2 = \varphi_2$, with details shown in S10(f) of Supplemental Materials [identical to FIG. S9(f)].

Obviously, the pair of EL3s protects the same number of edge states as that protected by both the NL and the NIL, because the pair of EL3s is topologically equivalent to the NL and NIL together. This equivalence relation is independent of PBC or OBC, despite the existence of the skin effect. This can be demonstrated by the identical eigenframe evolution process along GBZs and BZs (see Section 3 of Supplemental Materials), guaranteeing that their corresponding 1D periodic systems also have the same number of edge states (two pairs). As the BZ loop is deformed from $l_a$ to $l_b$ without encountering degeneracy lines (encountering ESs is allowed) the eigenvalues of the two pairs of edge states initially coalesce on the real axis before bifurcating in their imaginary parts, as demonstrated in the video. Although in both systems, the eigenvalues of edge states seem to reside inside the gap between $j_1$ and the merged continuum of $j_{2,3}$ (FIG. 3c,h), the edge states are independent of the gap. As both $l_a$ and $l_b$ can be deformed to intersect the ES formed by $j_1$ and $j_2$ (see $l_c$ in FIG. 3k). In this case, all the continuums are overlapped (FIG. 3l), but the two pairs of edge states remain robust against this gap closing (FIG. 3m), which well illustrates FIG. 1f. We thus conclude that the topology remains intact if the BZ does not encounter the three types of singularities lines. In a noteworthy observation, a 1D system with its BZ encircling a single EL3 lacks topological edge states. This is attributed to the unquantized accumulated angles of eigenstates resulting from an order change of the three eigenvalues along the BZ, thereby preventing the presence of edge modes (see results in Section 5 of the Supplemental Materials). Additionally, there exists another topologically equivalent relation between the double EL3 together with the NIL and the NL alone. Relevant discussions are shown in Section 3 (FIG. S6-S7) of the Supplemental Materials.

To summarize, we conducted both theoretical and experimental investigations into the existence of a whole new form of edge states in non-Hermitian topological gapless phases. The edge states in this context goes beyond the established paradigms of typical band gaps, as the edge states remain robust even when the line gap closes at EPs. Since conventional line-gap-based topological invariants cannot be defined when the gap closes, we proposed a theoretical framework that incorporates eigen-frame rotation and deformation, along with non-Bloch band theory, to address the challenges posed by inapplicability of band-gap theories. This framework effectively deals with the issue of gap closing at EPs in both the BZ and the GBZ. Additionally, we demonstrated that the topological connection between the three different forms of degeneracy lines in the swallowtail shape leads to interconnections in the number of edge states in systems with 1D BZs encircling distinct degeneracy lines. While we specifically chose a system that exhibits swallowtail features due to its rich hypersurface singularities, this novel form of topological edge states is not limited to the swallowtail and can be readily expanded to other hypersurface singularities that are prevalent in non-Hermitian systems with *PT* symmetry. Our work constitutes a major advance in extending the topological edge states from gapped phases to gapless phases. The study also lays the groundwork for investigating the relationships between high-order topological corner modes or hinge modes and such gapless phases.

### ACKNOWLEDGEMENTS

This work is supported by RGC Hong Kong through grants 16307621, 16307821, 16310422 and AoE/P-502/20. We acknowledge Prof. S. Liu for helpful comments in experimental implementation.

**End Matter**

**1. Dispersions nearby degeneracy lines:** The band dispersions nearby the three types of degeneracy lines are shown in FIG. 4(a-c), corresponding to the pair of EL3s, the NIL and the NL, respectively [on the cross sections 1-3 enclosed by the dashed blocks in FIG. 2(a)]. The EL3s are identified as three-fold defective degeneracy hosting only one eigenstate. The NIL manifests as a self-linear crossing of the ES formed by the 2nd and 3rd bands, and the NL presents as a linear cross between the 1st and 2nd bands in the exact phase, thus both the NIL and NL are non-defective degeneracies with complete eigenstates.

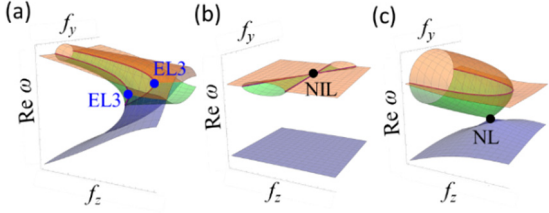

FIG. 4. Band dispersions nearby the three types of degeneracy lines. (a-c) correspond to the nearby areas [on the cross sections 1-3 in FIG. 2(a)] of the EL3s, the NIL and the NL, respectively.

**2. Edge states protected by a single NL or a single NIL:** We have determined that a 1D system with BZ encircling the NL and the NIL hosts two pairs of edge states, here we investigate the edge states associated with a single NL or a single NIL. We begin by examining the simplest degeneracy line, the NL, which is isolated from ESs. Since our considered swallowtail includes several ESs in the **f**-space [FIG. 2(a)], loops encircling NL can intersect the ESs, as illustrated by the loop $l_1$ (red ellipse) in FIG. 5(a). The cut plane on which $l_1$ resides is shown by the inset. In this context, our system exhibits a gapless band structure with two EPs on the BZ, as shown by the dispersion (real part of eigenvalues) in FIG. 5(b), where the $j_2$ and $j_3$ undergo gap closing (dashed yellow lines), and both $j_2$ and $j_3$ remain isolated from the $j_1$. Figure 5(c) shows the complex eigen-spectrum under PBC (olive symbols) and OBC (brown symbols). The system is found to exhibit one pair of edge states at the two boundaries (labelled 1 and 2), as indicated by the field distribution in FIG. 5(d). We can visualize the topological nontriviality by observing the accumulated rotation angle of eigenframe within the rotation and deformation process along GBZ. It is shown that the accumulated rotation angle of $\varphi_3$ is zero, and this accumulated angle results in $\hat{O}\varphi_3 = \varphi_3$. Both $\varphi_1$ and $\varphi_2$ accumulate a quantized $\pi$ rotation angle [see FIG. 5(d)], which maps $\hat{O}\varphi_1 = -\varphi_1$, $\hat{O}\varphi_2 = -\varphi_2$ [see FIG. S6(f) in Supplemental Materials]. The evolution of the eigenstates in BZ gives the same results, and we provide comparison in Section 3 of Supplemental Materials to avoid repetitiveness. This consistency allows us to adopt the same criterion used for PBC to identify bulk topology under OBC. Additionally, each type of eigen-frame evolution along a 1D loop uniquely indicates the type of singular lines enclosed by the loop. This allows for establishing a clear correspondence between the edge states and the singular lines in the 3D **f**-space.

Considering the case of NL, it has been revealed that the edge states are robust against line gap closing between $j_2$ and $j_3$. This observation may not be too surprising, as there still exists a gap between the lowest band $j_1$ and the upper bands $j_{2,3}$, and only one pair of edge states reside inside that gap [FIG. 5(c)]. The case for NIL will be fundamentally different. Since the NIL resides on the intersection of ESs, a loop encircling the NIL will inevitably cut through the ESs four times [see $l_2$ in FIG. 5(f)]. By mapping the $l_2$ loop to the BZ of a 1D periodic system, there will be four EPs on the BZ, and thus $j_2$ and $j_3$ undergo gap closing four times [see FIG. 5(g)], which eliminates the gap between $j_2$ and $j_3$. Despite this phase is essentially gapless, a pair of edge states (labelled by 1 and 2) with eigenvalues residing on the real axis can still be observed, as indicated by the experimental and theoretical results in FIG. 5(h). As a notable feature in the projection bands, the eigenvalues of the edge modes are surrounded by the eigenvalues of the continuous modes formed by $j_{2,3}$, instead of situated within any gap. The experimental validity of the field confinement for the two edge states is depicted in FIG. 5(i). Obviously, the emergence of edge states is not due to gap topology but rather stems from the gapless topology contributed by the NIL. This can be demonstrated by the eigenframe evolution in GBZ. As shown in the left panel of FIG. 5(j), $\varphi_2$ (blue) and $\varphi_3$ (red), both accumulate quantized $\pi$ rotation angles. Consequently, $\varphi_2$ and $\varphi_3$ both evolve to the antipodal points of their initial states $\hat{O}\varphi_2 = -\varphi_2$, $\hat{O}\varphi_3 = -\varphi_3$. Meanwhile, the accumulated angle of $\varphi_1$ is zero, and thus $\varphi_1$ evolves to its initial state ($\hat{O}\varphi_1 = \varphi_1$) [see FIG.

S8(f) in Supplemental Materials]. Hence, the topology of the NIL, which is a linear cross between bands $j_2$ and $j_3$, is encoded by the quantized accumulated angles of their corresponding eigenstates $\varphi_2$ and $\varphi_3$, while the trivial angle for $\varphi_1$ has no significance.

It is consistent with the observation that the eigenvalue of edge states protected by the topology of the NIL is surrounded by the continuum of $j_{2,3}$ (as $j_2$ and $j_3$ are degenerate), rather than in the line gap between $j_1$ and $j_{2,3}$.

Obviously, a 1D system with BZ enclosing a single NL or a single NIL shows two-band topological behaviors, because both singular lines are linear crosses between two bands (NL: formed by $j_1$ and $j_2$, NIL: formed by $j_2$ and $j_3$). This can also be manifested by quantized accumulated angles via eigenframe deformation and rotation processes (NL: $\pi$ for $\varphi_1$ and $\varphi_2$; NIL: $\pi$ for $\varphi_2$ and $\varphi_3$). Therefore, the assembling of them shows a multiband topological behavior ($\pi$ for $\varphi_1$ and $\varphi_3$), which is explicit in FIG. 3(a-e). The number of edge states (two pairs) in a 1D system with BZ enclosing the NL and NIL together is thus understandable, as the evolution process [FIG. 3(e)] aggregates the processes in FIG. 5(e,j).

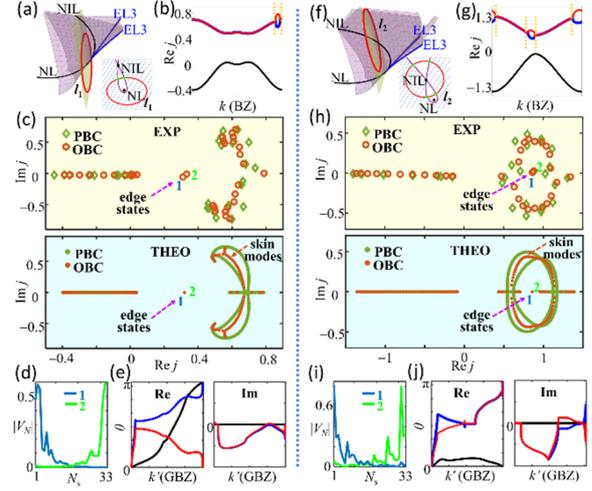

FIG. 5. Experimental observation of topological edge states in 1D periodic systems with BZs encircling NL and NIL. (a) BZ of the periodic system (Red loop, $l_1$) in 3D **f**-space that encircles the NL. The cross-section plane that the BZ resides (yellow plane) is shown by the inset. (b) Real part of band dispersion in BZ of the system (i.e., on $l_1$). (c) Projection bands on the complex plane of eigenvalues of the 1D system under PBC (olive colored) and OBC (brown colored) that are obtained experimentally (upper panel) and theoretically (lower). One pair of topological edge modes are labelled by the purple arrows, with eigenvalues distinctly away from the continuous modes. The two edge states are labelled with 1 and 2 in the figures. (d) Experimental results of the field distributions (voltage response) on each lattice site for the two edge states 1 and 2 in panel (c). (e) Topological characterization with the eigenstates evolution on the GBZ. (f-j) Similar to panels (a-e) but for the 1D system with BZ encircling the NIL.